%% file: ms.tex
\documentclass{latex/webofc}
\input{latex/packages.tex}

\input{latex/commands.tex}

\begin{document}

%
\title{Bayesian Methodologies with \texttt{pyhf}}

\author{\firstname{Matthew} \lastname{Feickert}\inst{1}\fnsep\thanks{\email{matthew.feickert@cern.ch}} \and
 \firstname{Lukas} \lastname{Heinrich}\inst{2}\fnsep\thanks{\email{lukas.heinrich@cern.ch}} \and
 \firstname{Malin} \lastname{Horstmann}\inst{2}\fnsep\thanks{\email{malin.elisabeth.horstmann@cern.ch}}
}

\institute{University of Wisconsin-Madison, Madison, Wisconsin, USA
 \and
 Technical University of Munich, Munich, Germany
}

\abstract{%
 \input{src/abstract}
}
\maketitle

\input{src/introduction}
\input{src/BayesForPyhf}
\input{src/Workflow}
\input{src/conclusions}

\section{Acknowledgements}\label{sec:acknowledgements}

MH and LH are supported by the Excellence Cluster ORIGINS, which is funded by the Deutsche Forschungsgemeinschaft (DFG, German Research Foundation) under Germany’s Excellence Strategy - EXC-2094-390783311.
MF is supported by the U.S. National Science Foundation (NSF) under Cooperative Agreement OAC-1836650 (IRIS-HEP).

\bibliography{bib/ref}

\end{document}

%% file: latex/commands.tex
\newcommand{\pyhf}{\texttt{pyhf}}

\newcommand*{\TeV}{\ensuremath{\text{Te\kern -0.1em V}}}
\newcommand*{\GeV}{\ensuremath{\text{Ge\kern -0.1em V}}}
\newcommand*{\MeV}{\ensuremath{\text{Me\kern -0.1em V}}}
\newcommand*{\keV}{\ensuremath{\text{ke\kern -0.1em V}}}
\newcommand*{\eV}{\ensuremath{\text{e\kern -0.1em V}}}

%% file: src/abstract.tex
\texttt{bayesian\_pyhf} is a Python package that allows for the parallel Bayesian and frequentist evaluation of multi-channel binned statistical models. The Python library \texttt{pyhf} is used to build such models according to the \texttt{HistFactory} framework and already includes many frequentist inference methodologies. The \texttt{pyhf}-built models are then used as data-generating model for Bayesian inference and evaluated with the Python library \texttt{PyMC}. Based on Monte Carlo Chain Methods, \texttt{PyMC} allows for Bayesian modelling and together with the \texttt{arviz} library offers a wide range of Bayesian analysis tools.

%% file: src/introduction.tex
\section{Introduction}\label{sec:introduction}

The evaluation of High Energy Physics measurements depend on the comparison with theoretical predictions. The phenomenology of the observation is represented by a statistical model $p(x | \theta)$, i.e. the probability distribution of data $x$ for specific theory parameters $\theta$.
Given actual observations $x$, the likelihood $\mathcal{L}(\theta)$ can then be interpreted as $p(x | \theta)$ with fixed $x$ and is a measure of the compatibility between the observed data $x$ and the theory prediction depending on $\theta$.\\ \\
Based on $p(x | \theta)$ there are different approaches to evaluating the model. In the frequentist setting inference methodologies include maximum-likelihood point estimation, hypothesis tests and confidence interval estimation. Applying Bayesian statistics to $p(x | \theta)$ is a different approach where the likelihood is used to update a prior belief $p(\theta)$ to a posterior belief $p(\theta|x)$ about the probable values of the model parameters $\theta$. \\
In particle physics statistical models are often represented based on templates such as \texttt{HistFactory}. \texttt{HistFactory} is a mathematical framework for building statistical models of binned analyses across different channels, see Sec.~\ref{subsec:HFandpyhf}. \texttt{RooFit}~\cite{root} is a framework that already allows for Bayesian inference for \texttt{HistFactory} models, its range of application though is limited by the lack of the implementation of gradients and availability of advanced diagnostics for Bayesian inference results due to the historical focus on frequentist inference in HEP. An example for a library that allows for advanced Bayesian inference for particle and astro-physics is \texttt{BAT.jl}~\cite{Schulz:2021BAT} but tools to construct \texttt{HistFactory} models within Julia are not yet readily available~\footnote{The LiteHF.jl~\cite{LiteHF} package is working towards HistFactory within a Julia context.}. \\
\texttt{pyhf} is a Python library that implements the \texttt{HistFactory} template and already allows for frequentist inference~\cite{pyhf, pyhf_joss}. It is the aim of this work to utilize the Python library \texttt{PyMC} and the automatic differentiation capabilities of \texttt{pyhf} to enable advanced Bayesian analysis for \texttt{HistFactory} models.

%% file: src/BayesForPyhf.tex
\section{Bayesian Statistics for HistFactory Models}\label{BayesForpyhf}

\subsection{\texttt{HistFactory} Models and \texttt{pyhf}}\label{subsec:HFandpyhf}

In the \texttt{HistFactory} template, the expected event rates $\nu$ are dependent on two sets of parameters, free parameters $\eta$ and constraint parameters $\chi$. In contrast to the free parameters, the $\chi$ are constrained by external data, whose impact has to be considered when building the statistical model. This can be done by assuming auxiliary measurements with observations $a_{\chi}$ for each parameter $\chi$. Each auxiliary measurement then corresponds to a constraint term  $c_{\chi}$ which is added to the likelihood and controls the compatibility of the value of the constraint parameter $\chi$ with their corresponding auxiliary measurements $a_{\chi}$~\cite{pyhf, pyhf_joss, Cranmer:1456844}. For simplicity, the model for these auxiliary measurements are either Gaussian or Poisson distributions. \\
Taking the constraints into account, the resulting statistical model for event rates $n$ and auxiliary measurements $a$ is then given by~\cite{pyhf, pyhf_joss}:

    \begin{align}
    \begin{split}
        p(x | \theta ) &= p_{\text{main}} (x_{\text{main}}| \theta ) p_{\text{aux}} (x_{\text{aux}}| \theta) \\
         &= p( n, a | \eta, \chi) = \prod_{c \in \text{channels}}  \prod_{b \in \text{bins}} \text{Poiss}(n_{cb} | \nu_{cb}(\eta, \chi)) \prod_{\chi}c_{\chi}(a_{\chi} | \chi),
    \end{split}
    \end{align}

\noindent where $p_{\text{main}}$ ($p_{\text{aux}}$) and $x_{\text{main}}$ ($x_{\text{main}}$) describe the actual and auxiliary statistical model and observations for parameters $\theta$. \\ \\
\noindent The pure-Python library \texttt{pyhf} implements the \texttt{HistFactory} formalism for the analysis of multi-channel binned statistical models~\cite{pyhf, pyhf_joss}. Statistical models can be stored as pure JSON files, allowing for integration with other statistics libraries. The numeric backend that $\pyhf$ uses is flexible, allowing for auto-differentiable tensor-backends.

\subsection{Bayes' Theorem}
Bayesian inference is governed by Bayes' theorem~\cite{ConjPriorsBerkeley}:
    \begin{align} \label{Bayes}
        p(\eta, \chi \vert x, a) = \frac{p(x, a\vert \eta, \chi) p(\eta, \chi)}{p(x, a)}.
    \end{align}
 \noindent It describes the updating of a prior probability distribution $p(\eta, \chi)$ to a posterior distribution $p(\eta, \chi \vert x)$ by multiplication with a data-generating model $p(x \vert \eta, \chi)$. $\eta, \chi$ are parameters of interest (POI) and constraint parameters respectively and $x, a$ observations and auxiliary measurements and evidence $p(x, a)$. While a closed form solution of Eq.~\eqref{Bayes} is intractable due to the evidence, approximate solutions are viable via sampling methods (such as MCMC~\cite{PyMC}) as only a tractable joint likelihood $p(x, a \vert \eta, \chi)p(\eta, \chi)$ is required.

\subsection{\texttt{PyMC}}
\texttt{PyMC} is a Python library for building Bayesian models and already includes a wide range of cross-checks and plotting functions through its \texttt{arviz}-backend~\cite{PyMC, arviz}. \\
\noindent The statistical models are evaluated using Monte Carlo chain methods (MCMC), where the posterior is represented by sampling from the prior distribution steered by the likelihood.
\noindent \texttt{PyMC} also allows for the implementation of external models, which makes it suitable for performing Bayesian inference with \texttt{pyhf}-based \texttt{HistFactory} models. \\
Within \texttt{PyMC} a whole set of MCMC techniques is available, e.g. prior and posterior sampling or predictive sampling. The returned objects are \texttt{arviz.InferenceData} containers, for which again the whole set of analytic tools provided by the \texttt{arviz} library are available~\cite{arviz}.

\subsection{Prior Constraints from Auxiliary Measurements and Ur-Priors}

In order to get a sampling representation of the posterior using MCMC methods the prior distribution and the \texttt{HistFactory} models are needed, i.e following Eq.~\eqref{Bayes}:
    \begin{align}
        p(\eta, \chi \vert x, a) \approx p(x, a \vert \eta, \chi)p(\eta, \chi).
    \end{align}
\noindent While $p(x, a \vert \eta, \chi)$ can be build using \texttt{pyhf}, the prior beliefs of the value of the parameters still have to be quantified. It would be possible to treat $\eta$ and $\chi$ equally, i.e. to determine some ur-prior $p(\eta, \chi)$ by hand and update with $x$ and $a$ in parallel. Ur-priors describe the belief about the parameter value before taking the observations into account. \\
The approach followed in this work however relies on a different treatment of the constraint priors $p(\chi)$. It is based on separating the auxiliary observations $a$ from the main inference step and using it instead in an initial inference step. In this step, Bayes' theorem is used to update ur-priors $p_\text{ur}(\chi)$ with the auxiliary measurements $a$, thus incorporating the information gained from the auxiliary measurements $a$, see Eq.~\eqref{initialInference}.
    \begin{align} \label{initialInference}
	p(\chi \vert a) \approx p(a \vert \chi) p_{\text{ur}}(\chi)
    \end{align}

\noindent The posteriors $p(\chi \vert a)$ from Eq.~\eqref{initialInference} can then be used as prior belief in the main inference step. \\
This approach is useful, as the number of constraint measurements can get arbitrarily high which would imply high computational cost when updating $p(\eta)$ and $p(\chi)$ together. Splitting the auxiliary update of the constraint parameters into this initial step solves this issue based on the concept of conjugate priors. This concept dictates that for given sets of distributions for the priors and the data-generating model, the posterior distribution is of the same distribution family as the prior and can be given in closed form --- the priors and posteriors are then conjugate. Due to the limited possibilities for the auxiliary measurements (which can be either Gaussian or Poisson distributed), this concept is viable for the constraint parameters with corresponding Gaussian and Gamma distributed ur-priors, see Table~\ref{ConjugatePriors}~\cite{ConjPriorsBerkeley}. The freedom to choose these compatible ur-priors is justified as the priors will be dominated by the auxiliary measurements. Indeed, in the limit of very vague ur-priors the priors are completely dominated by the auxiliary measurements $a$:
    \begin{align}
    \begin{split}
        \sigma_{\text{ur}} \gg \sigma_{\text{aux}} \quad &\longrightarrow \quad  \mu' \rightarrow a, \quad \sigma' \rightarrow \sigma_{\text{aux}}, \\
        \alpha_{\text{ur}}, \beta_{\text{ur}} \rightarrow 0 \quad &\longrightarrow \quad \alpha', \beta' \rightarrow a.
    \end{split}
    \end{align}
As the impact of the auxiliary measurements can now be implemented in closed form, no sampling is needed for the implementation of the knowledge gained from the auxiliary measurements $a$.

    \begin{center}
        \begin{tabular}
        {| c || c | c|}
         \hline
          Posterior & Data-Gen. Model & Ur-Prior \\
         \hline
         \hline
         $\mathcal{N}\left( \chi | \mu',~\sigma'\right)$ &$ \mathcal{N}\left( a | \chi,~\sigma_{\text{aux}}\right)$ & $\mathcal{N}\left( \chi | \mu_{\text{ur}},~\sigma_{\text{ur}}\right)$ \\
        \hline
        $\Gamma\left(\chi |\alpha', \beta'\right)$ & $\text{Poiss}\left( a | \chi \right)$ & $\Gamma\left(\chi |\alpha_{\text{ur}}, \beta_{\text{ur}} \right)$ \\
        \hline
        \end{tabular}
    \captionof{table}{Conjugate priors for Poisson and Normal distributed auxiliary measurements $a$~\cite{ConjPriorsBerkeley, Murphy2007}. \label{ConjugatePriors}}
    \end{center}

\noindent While Table~\ref{ConjugatePriors} fixes the general structure of the posterior distribution (to be used as priors in the main inference), the hyperparameters governing these still have to be determined. \\
For a single auxiliary measurement $a$, the hyperparameters for the Gaussian posteriors of $\chi$ for some given ur-hyperparameters $\mu_{\text{ur}}, \sigma_{\text{ur}}$ follow~\cite{Murphy2007}:

    \begin{align} \label{BayesConj}
        \mu'= \frac{\sigma_{\text{aux}}^2 \sigma_{\text{ur}}^2}{\sigma_{\text{aux}}^2 + \sigma_{\text{ur}}^2} \left( \frac{\mu_{\text{ur}}}{\sigma_{\text{ur}}^2} + \frac{a}{\sigma_{\text{aux}}^2}\right) , \quad \sigma' = \frac{\sigma_{\text{aux}}^2 \sigma_{\text{ur}}^2}{\sigma_{\text{aux}}^2 + \sigma_{\text{ur}}^2},
    \end{align}

\noindent The hyperparameters describing the posterior Gamma distribution for a single auxiliary observation $a$ can for some given ur-hyperparameters $\alpha_{\text{ur}}, \beta_{\text{ur}}$ be derived as:
    \begin{align} \label{PoissonUpdating}
        \alpha' = \alpha_{\text{ur}} + a, \quad \beta' = \beta_{\text{ur}} + 1,
    \end{align}
\noindent In contrast to the Gaussian constraints, the implementation of the Poisson constraints in \texttt{pyhf} comes with the following change of variable:
    \begin{align} \label{}
        \chi  \rightarrow \gamma = \frac{\chi}{a}.
    \end{align}
\noindent Using:
    \begin{align}
        p(\chi) \mathrm{d} \chi = p(\gamma) \mathrm{d} \gamma
    \end{align}
\noindent the posterior Gamma distribution over $\gamma$ is then derived as:
    \begin{align}
    \begin{split}
        p(\gamma | a ) &\approx \Gamma(\chi | \alpha', \beta') \frac{\mathrm{d} \chi}{\mathrm{d} \gamma} = a \frac{\chi^{\alpha'-1}}{\Gamma(\alpha')} \text{e}^{-\beta' a \chi} \\
        &\approx \Gamma(\gamma | \alpha', a\beta') = \Gamma(\gamma | \alpha_{\text{ur}} + a, a(\beta_{\text{ur}} + 1)).
    \end{split}
    \end{align}

\noindent Applying the methods described above to Eq.~\eqref{Bayes}, the final form of Bayes' theorem used in this work then reads:
    \begin{align} \label{BayesConj_final}
        p(\eta, \chi \vert x, a) \propto p(x\vert \eta, \chi) p(\eta) p(\chi | a).
    \end{align}

\subsection{Hamiltonian Monte Carlo Sampling}
\begin{figure}
        \centering
        \includegraphics[width=10cm]{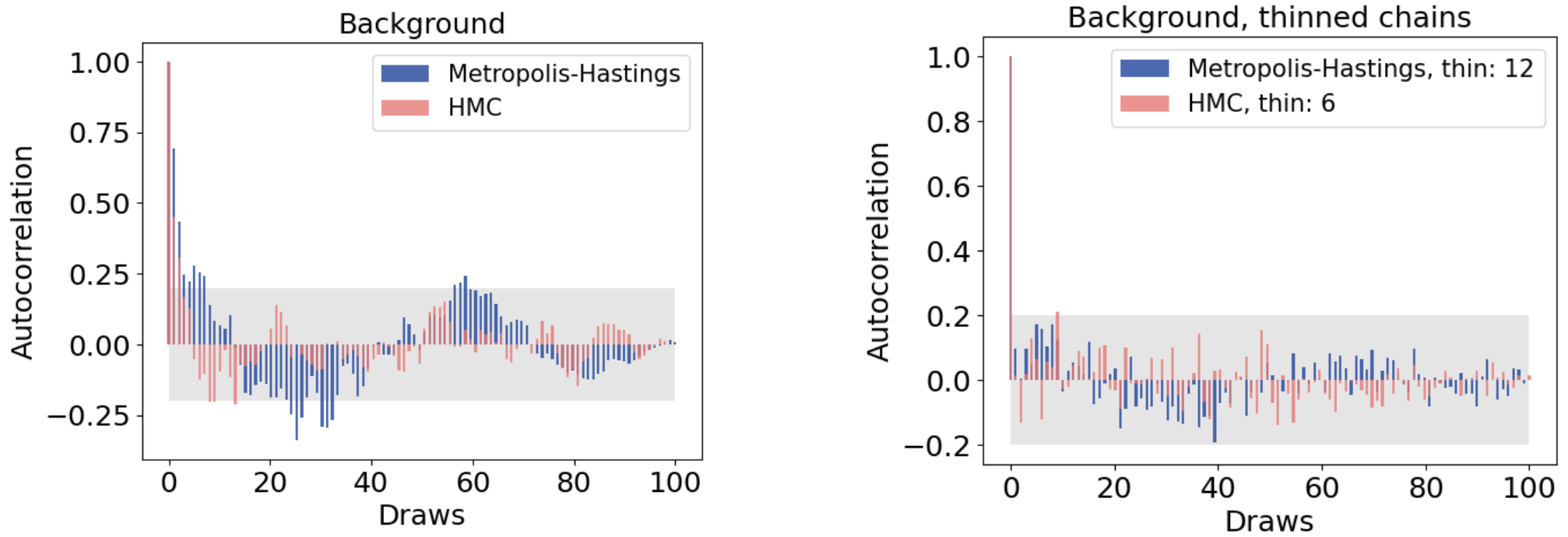}
        \centering
        \caption{Autocorrelation length for the background parameter for HMC and Metropolis-Hastings steps. The left plot shows the original chains, the right the thinned chains. The shaded band indicates the acceptable length.}
        \label{autocorr}
    \end{figure}

\texttt{pyhf} supports auto-differentiation via its \texttt{jax, torch} and \texttt{tensorflow} backends~\cite{pyhf, pyhf_joss, jax2018github, tensorflow2015-whitepaper, paszke2017automatic}. Therefore, the Hamiltonian Monte Carlo (HMC) step method provided by \texttt{PyMC} is viable for analysing \texttt{pyhf} \texttt{HistFactory} models. Details on the HMC sampling method can be found in~\cite{vishnoi2021introduction}. For this work, it is sufficient to point out that HMC relies on the derivatives of the model --- and while this comes with a computational cost, the quality of the drawn samples should be higher compared to other step methods, such as Metropolis-Hastings~\cite{Metropolis1953}. \\ \\
\noindent The quality of MCMC chains can be measured using the autocorrelation length, i.e. the correlation between subsequent samples. Independent samples are necessary to fully express the parameter space. In order to reduce the autocorrelation length, thinning can be applied. In thinned chains, only every $n$th sample is kept in the final chains~\cite{hoyer2017xarray}. Fig.~\ref{autocorr} visualises how the Metropolis-Hastings chains have to be thinned twice as much ($n=12$) compared to the HMC chains ($n=6$) in order to keep the autocorrelation within an acceptable range (see Sec.~\ref{BaysWF} for the model used). Accordingly, in order to produce sampling chains of the same magnitude, the computational cost of the gradient calculation can seen as substituting the cost for drawing twice as many samples for Metropolis-Hastings steps.

%% file: src/Workflow.tex
\section{A Bayesian Workflow}\label{BaysWF}

For testing and evaluating the Bayesian inference model we follow Ref.~\cite{Betancourt2020} and present the selected steps below. \\

\noindent We demonstrate the inference methods derived in Sec.~\ref{BayesForpyhf}, i.e. building a statistical model using \texttt{pyhf} and then evaluating it using the whole range of inference techniques provided by \texttt{PyMC} and \texttt{arviz}, using a simple model. This model has three bins and one signal strength parameter $\eta$ (the POI of the model) and one correlated background parameter $\chi$ (constrained by a Normal-distributed auxiliary measurement) with ur-hyperparameters $\mu_{\text{ur}}, \sigma_{\text{ur}} = 0, 2$. The event counts $n_i$ for each bin $i$ and a signal $s_i$ and background $b_i$ can be calculated as:
    \begin{align} \label{binCounts}
        n_i = \eta s_i + \chi b_i.
    \end{align}

\begin{figure} 
        \centering
             \begin{subfigure}[b]{0.35\textwidth}
                 \centering
                 \includegraphics[width=\textwidth]{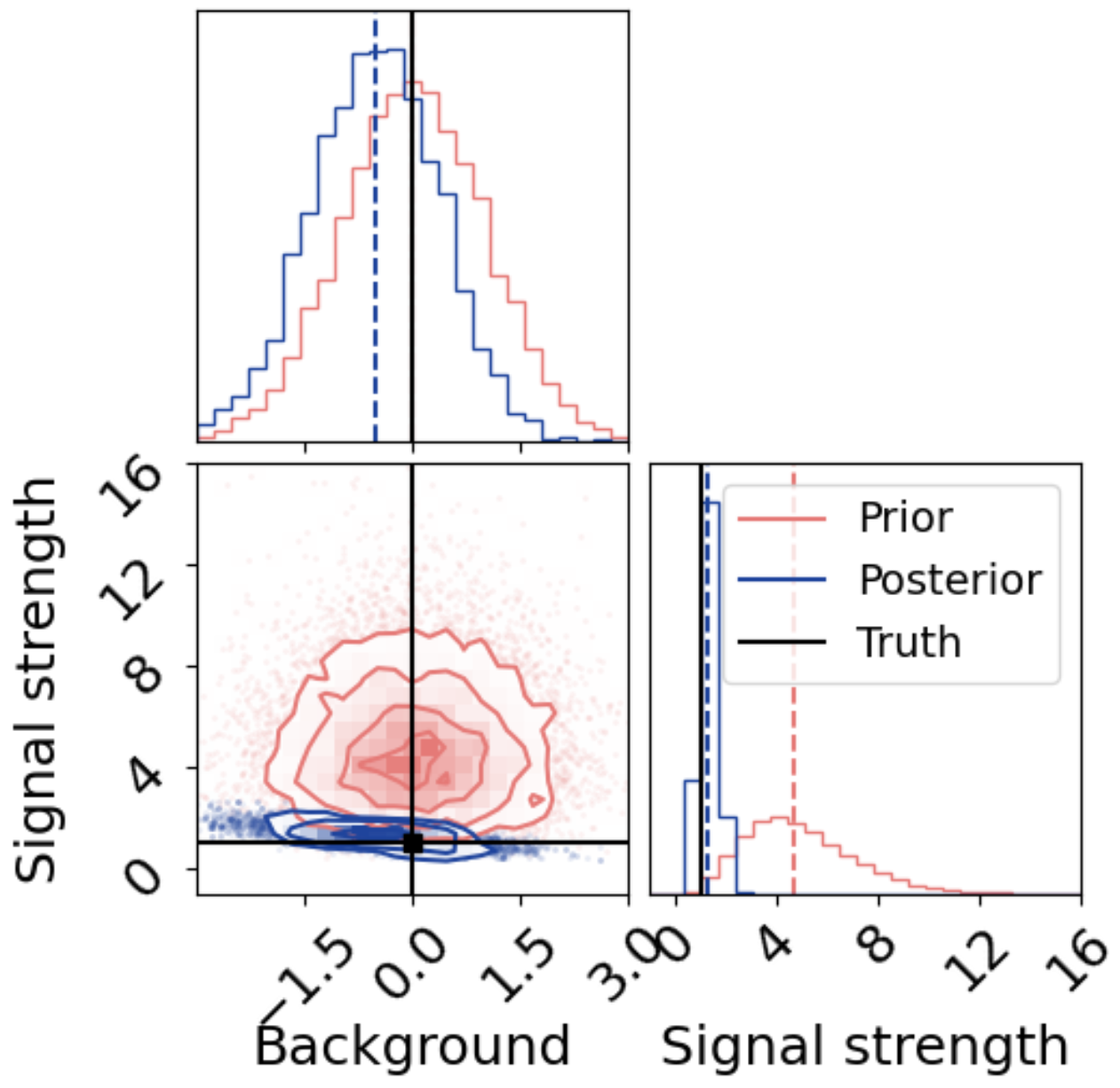}
                 \caption{Correlation for signal and background distribution. Indicated in black is the underlying truth.}
                 \label{corner}
             \end{subfigure}
        \hspace{0.2\textwidth}
             \begin{subfigure}[b]{0.3\textwidth}
                 \centering
                 \includegraphics[width=\textwidth]{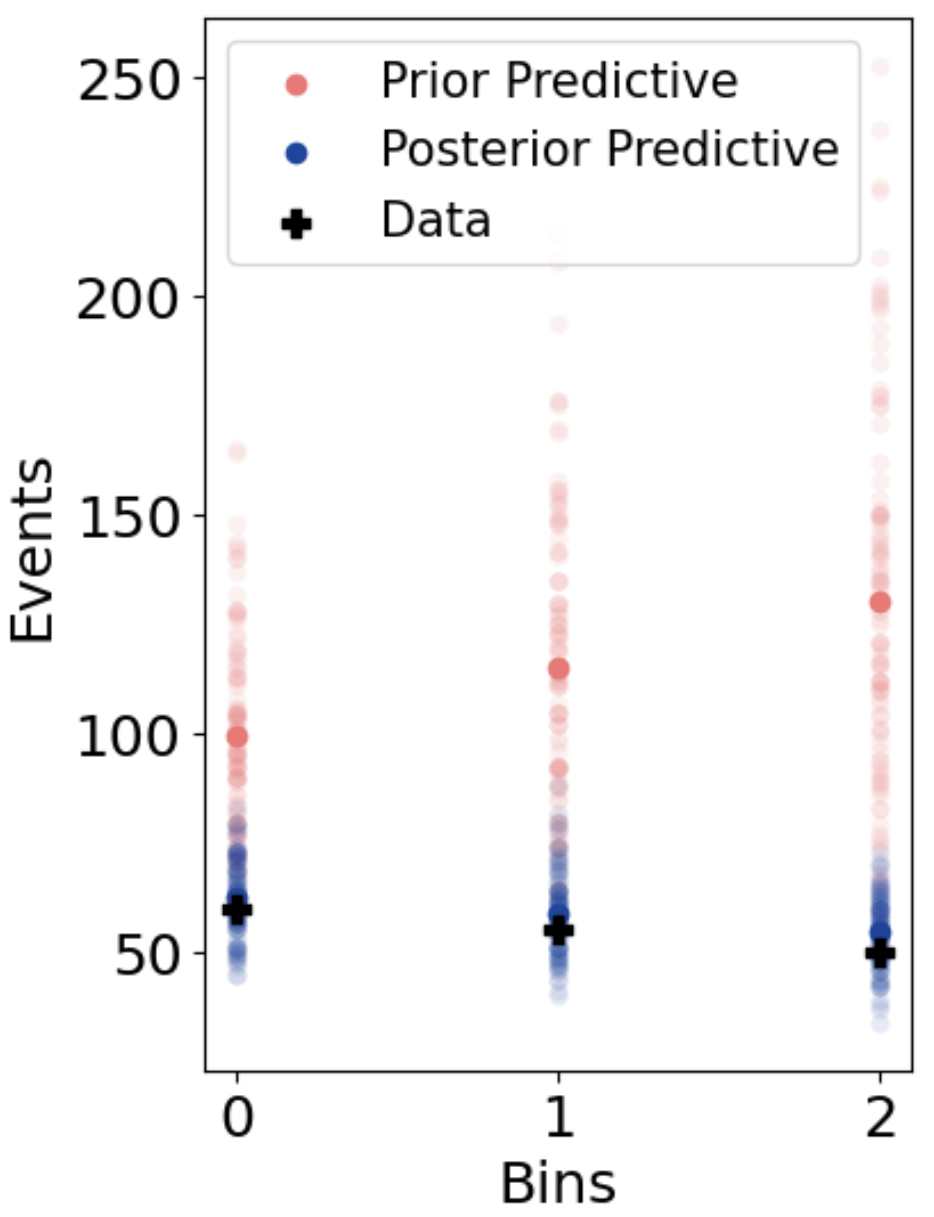}
                 \caption{Prior and posterior predictive. Indicated in black is the observed data.}
                 \label{ppc}
             \end{subfigure}
        \caption{Comparing prior and posterior distribution for the parameters (\ref{corner}) and predictions for the observations (\ref{ppc}).}
        \label{ppc_corner}
    \end{figure}

\begin{figure} 
        \centering
        \includegraphics[width=10cm]{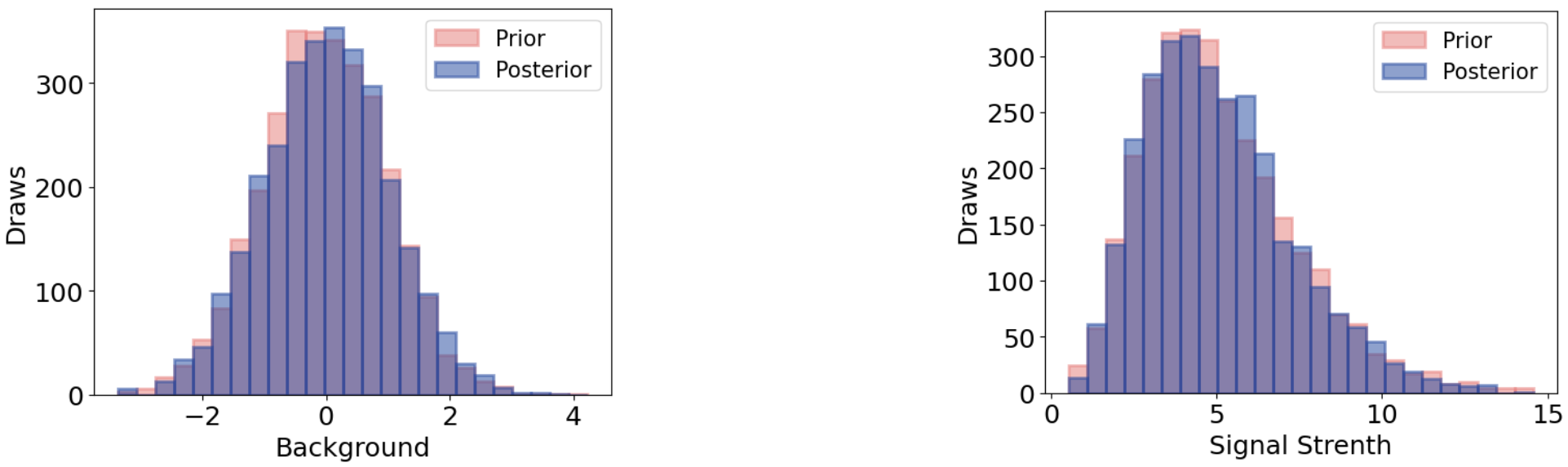}
        \centering
        \caption{Calibration check for the signal POI and the background using 3000 pseudo-observations drawn from the prior predictive.}
        \label{calibrationPlot}
    \end{figure}

\noindent The main result of Bayesian inference are the posterior parameter distributions, which can be used to predict observations (predictives). These results are visualized in Fig.~\ref{ppc_corner}.
\noindent A possible next check is a calibration check, which tests the computational faithfulness of the inference, i.e. whether the distribution of posterior samples can capture a distribution of pseudo-data observations. In particular, if a set of pseudo-observations $x$ is sampled from the prior predictive, the resulting distribution of posteriors should approximate the prior distribution, see Eq.~\ref{calibration}.
    \begin{align} \label{calibration}
        p\left(\eta, \chi\right) \overset{!} \approx \int dxd\eta' d\chi' ~  p\left( \eta, \chi | x \right) ~ p\left(y|\eta', \chi'\right)
    \end{align}
\noindent In Fig.~\ref{calibrationPlot} this is visualised for the simple model introduced above.

%% file: src/conclusions.tex
\section{Conclusions}\label{sec:conclusions}

The methods presented above are implemented in the Python package \texttt{bayesian\_pyhf}~\cite{BayesianPyhf}.
This software package enables the parallel Bayesian and frequentist analysis of multi-channel binned models within the single software framework \texttt{pyhf}.
The current interface of the package \texttt{bayesian\_pyhf} is demonstrated in Listing~\ref{lst:pymc_example}.
Further enhancements regarding the user interface and stability with respect to multi-chain sampling are ongoing.
A full integration in the \texttt{pyhf} library is also planned.

\begin{listing}
\begin{minted}{python}
with infer.model(model, unconstr_priors, data):
    post_data = pymc.sample(draws=10_000, chains=1)
    post_pred = pymc.sample_posterior_predictive(post_data)
    prior_pred = pymc.sample_prior_predictive(10_000)
\end{minted}
 \caption{Pseudo-code for evaluating \texttt{HistFactory} models (\texttt{model}) using \texttt{PyMC} given unconstrained parameters (\texttt{unconstr\_priors}) and observations (\texttt{data}). \texttt{post(prior)\_pred} are the posterior (prior) predictives and \texttt{post\_data} are the samples from the posterior distribution. Following the \texttt{PyMC} syntax~\cite{PyMC}, the \texttt{with} statement opens a context, that initializes the inference in a way that all actions within the block are interpreted with respect to the given model, data and priors. In addition, the methodologies regarding conjugate priors from Sec.~\ref{subsec:HFandpyhf} are applied under the hood, resulting in the constraint priors which are added to the model parameters for sampling.}
 \label{lst:pymc_example}
\end{listing}